\documentclass{elsarticle}
\usepackage{natbib}
\usepackage{rotating}

\usepackage{lineno}
\usepackage{graphicx}
\newcommand{\BibTeX}{ \textrm{B\kern-.05em\textsc{i\kern-.025em b}\kern-.08em
    T\kern-.1667em\lower.7ex\hbox{E}\kern-.125emX} }

\begin{document}

\begin{frontmatter}

% Title, authors and addresses

% use the thanksref command within \title, \author or \address for footnotes;
% use the corauthref command within \author for corresponding author footnotes;
% use the ead command for the email address,
% and the form \ead[url] for the home page:
% \title{Title\thanksref{label1}}
% \thanks[label1]{}
% \author{Name\corauthref{cor1}\thanksref{label2}}
% \ead{email address}
% \ead[url]{home page}
% \thanks[label2]{}
% \corauth[cor1]{}
% \address{Address\thanksref{label3}}
% \thanks[label3]{}

\title{Atmospheric waves and dynamics beneath Jupiter's clouds from radio
wavelength observations}

% use optional labels to link authors explicitly to addresses:
% \author[label1,label2]{}
% \address[label1]{}
% \address[label2]{}

\author[RGC]{Richard G. Cosentino} 
\author[BB]{Bryan Butler}
\author[BS]{Bob Sault} 
\author[RGC]{Raul Morales-Juberias}
\author[AS]{Amy Simon}
\author[IdP]{Imke de Pater}
%\author[PG]{Patrick Gaulme}

\address[RGC]{New Mexico Institute of Mining and Technology, Socorro NM 87801 (U.S.A.)}
\address[BB]{National Radio Astronomy Observatory, Socorro NM 87801 (U.S.A.)}
\address[BS]{Australia Telescope National Facility, CIRSO, Epping, NSW 2121 (Australia)}
\address[AS]{Solar System Exploration Div., NASA/GSFC, Greenbelt MD 20771 (U.S.A.)}
\address[IdP]{University of California, Berkeley CA 94720 (U.S.A.)}
%\address[PG]{New Mexico State University, Las Cruces NM 88003 (U.S.A.)}

%\begin{center}
%\scriptsize
%Copyright \copyright\ 2016, Richard G. Cosentino
%\end{center}

\begin{abstract}
%\begin{linenumbers}

We observed Jupiter at wavelengths near 2 cm with the Karl G. Jansky Very Large Array (VLA) in February 2015. These frequencies are mostly sensitive to variations in ammonia abundance and probe between $\sim0.5-2.0$ bars of pressure in Jupiter's atmosphere; within and below the visible cloud deck which has its base near 0.7 bars. The resultant observed data were projected into a cylindrical map of the planet with spatial resolution of $\sim$1500 km at the equator. We have examined the data for atmospheric waves and observed a prominent bright belt of radio hotspot features near 10$^\circ$N, likely connected to the same equatorial wave associated with the 5-micron hotspots. We conducted a passive tracer power spectral wave analysis for the entire map and latitude regions corresponding to eastward and westward jets and compare our results to previous studies. The power spectra analysis revealed that the atmosphere sampled in our observation (excluding the NEB region) is in a 2-D turbulent regime and its dynamics are predominately governed by the shallow water equations. The Great Red Spot (GRS) is also very prominent and has a noticeable meridional asymmetry and we compare it, and nearby storms, with optical images. We find that the meridional radio profile has a global north-south hemisphere distinction and find correlations of it to optical intensity banding and to shear zones of the zonal wind profile over select regions of latitude. Amateur optical images taken before and after our observation complemented the radio wavelength map to investigate dynamics of the equatorial region in Jupiter's atmosphere. We find that two radio hotspots at 2 cm are well correlated with optical plumes in the NEB, additionally revealing they are not the same 5 micron hotspot features correlated with optical dark patches between adjacent plumes. This analysis exploits the VLA's upgraded sensitivity and explores the opportunities now possible when studying gas giants, especially atmospheric dynamics of layers beneath upper level clouds. 
%\end{linenumbers}
\end{abstract}
 

\end{frontmatter}

\section{Introduction}
%\begin{linenumbers}
Understanding the dynamics of the giant planet atmospheres is a keystone to understanding the planets themselves, and also gives important clues to understanding extrasolar giant planets as well. In addition, comparative analysis of the dynamics in these atmospheres gives us information allowing for better understanding of Earth's atmosphere. We have much information on the upper cloud layers' dynamics for all of the giant planets, in particular Jupiter, from sensitive, high-resolution optical imaging with HST \citep{2009_AsayDavis, 2011_AsayDavis}, as well as multiple spacecraft visit flybys \citep{1979_Smith, 1979_Smith2, 1986_Limaye, 1989_limaye, 1996_Belton, 2003_Porco, 2007_Reuter}. However, how the dynamics of these upper regions couple to the dynamics of unseen regions beneath them and to the deeper atmosphere (referring to the depth of the water condensation level near 5 bars or greater), is still poorly understood, and yet is vital to understanding the planet as a whole \citep{1993_Gierasch}. Some models of Jupiter suggest that the upper layers of the atmosphere, like those observed at optical wavelengths, are connected via cylinders extending through the planet's interior and hence winds at depth are just as strong as those at the visible cloud layer \citep{1976_Busse}. Alternative models suggest that the coupling between the upper weather layer and the deep interior is weak, being forced by deep convection, and so the winds at depth are not strongly correlated with those at the visible cloud layer \citep{1969_Ingersoll}. The Galileo probe measured winds in a strong eastward jet that approached a constant speed from around 4 bars to a pressure up to 20 bars \citep{1998_Atkinson, 2005_Vasavada}. Modeling of cloud-level phenomena, as presented in \citet{2008_SanchezLavega} for example, match the Galileo probe observation for a different strong eastward jet but there is no clear consensus if this treatment of vertical wind shear is applicable globally or is specific for only these eastward jets \citep{2004_Ingersoll}.

Images of Jupiter at optical wavelengths probe the uppermost cloud layer in Jupiter's atmosphere - ammonia ice clouds with a base near 0.7 bar pressure; the exact value depends on abundance \citep{1969_Lewis, 2001_dePater, 2005_dePater}. Such observations are a great way to investigate the dynamics at the top of this cloud layer, since there is enough sensitivity and to make high-resolution images, shortly spaced in time to track small and large scale features. The high resolution ($\sim$0.1$^\circ$) maps produced from Hubble Space Telescope (HST) observations display fine structure in the upper cloud deck, such as small scale storms, eddies, and waves \citep{2012_SimonMiller, 2015_Simon}. Such features are also seen at even higher resolutions in spacecraft data \citep{2000_Ingersoll, 2000_Gierasch}. Some of these features resemble those seen in the Earth's atmosphere, where they are caused by convective processes; a similar origin is almost certainly true for Jupiter's atmosphere where there is evidence of lightning and scant water ice detections \citep{1999_Little, 2000_Gierasch, 2000_SimonMiller}.

While optical wavelength observations are extremely valuable, they cannot tell us about the dynamics beneath that uppermost cloud layer. For that, longer wavelengths, such as 5-micron infrared and radio, are required. Such observations can penetrate the uppermost cloud layers and probe the atmospheric winds and structure at pressures greater than 0.7 bars. Infrared observations near 5 microns probe pressures of $\sim5-7$ bars \citep{2015_Bjoraker}, and show prominent thermal structures; most notable are the infrared hotspots near 10$^\circ$N \citep{1996_Harrington}. The 5-micron hotspots have been correlated to holes seen in optical wavelength maps \citep{1996_Orton} (hence in the uppermost cloud layer), and radio emission at 2 cm in maps taken near the time of the Galileo Probe entry \citep{2004_Sault}. The correlated features in observations made at different wavelengths are thought to be the manifestation of a long lived equatorially trapped Rossby wave of wavenumber near $k=8$ \citep{1998_Ortiz, 2000_Showman, 2005_Friedson, 2013_Choi}. 

As an example of infrared observations being used to probe dynamics below the cloud layer, \citet{1996_Harrington} observed Jupiter at 4.9 microns to search for longitudinal wave structure. They did this via analysis of the thermal emission power spectrum in 1$^\circ$ averaged latitudinal bands. They detected the wave signature of the 5-micron hotspots, and also probed fine scale structure to a high wavenumber limit near $k\sim70$. This corresponds to a physical size slightly larger than 2.5$^\circ$ at the equator, where projection effects increase with latitude towards the poles and resolution worsens.

Radio wavelength observations can probe deeper still; at a wavelength of 20 cm, depths well over 10 bars are probed \citep{2001_dePater, 2005_dePater}. Wavelengths near 2 cm probe pressures just below the uppermost cloud deck; $\sim0.5-2.0$ bars \citep{2016_dePater}. It can be argued that observations at longer wavelengths are preferred, since they investigate greater depths. However, Jupiter has a strong magnetic field and emitted synchrotron radiation becomes a significant issue (contaminating the signal from the atmosphere) for Earth-based observations at longer wavelengths \citep{2003_dePater}. Observing at a wavelength of 2 cm is a good compromise of probing beneath the uppermost clouds while minimizing this contamination, since synchrotron is $<$ 3\% of the total emission at this wavelength \citep{2008_Kloosterman}. In addition, resolution worsens at longer wavelengths (for fixed observing telescope or synthesis array geometry), not allowing us to see finer scale features at these depths. Using the Very Large Array (VLA) to observe at 2 cm, one can obtain a fairly fine spatial resolution between ground based infrared facilities like IRTF ($\sim$2.5$^\circ$ near Jupiter's equator) and those of HST ($\sim$0.1$^\circ$). Radio wavelength observations at 2 cm with the VLA therefore, vertically connect the 2-4 bar atmosphere accessed at infrared wavelengths with coarse resolution to the visible cloud deck at high resolution by probing intermediate pressures with an intermediate resolution. Note that the Juno spacecraft will have a similar spatial resolution, but at much longer wavelengths; and using Juno spacecraft observations to constrain the deep dynamics is an important goal of that mission \citep{2012_Janssen, 2013_Liu}. Unfortunately, eventually the Juno science orbits will be separated by 14 days and view only narrow slices of Jupiter, which will not achieve good global context spatially or temporally. To address these shortcomings, Earth-based observations that can probe below the upper cloud layer are of utmost importance.

Recently, \citet{2016_dePater} published VLA observations spanning a wide range of frequencies. Some of their main findings can be summarized as follows: 1) They identified a radio-hot belt at the interface between the North Equatorial Belt (NEB) and the Equatorial Zone (NZ), consisting of relatively transparent regions, which might coincide with the 5-micron hot spots. Radiative transfer calculations show that the ammonia abundance (the main source of opacity at these wavelengths) in these regions is very low down to depths of at least $\sim8$ bars. 2) They detected ``plumes" containing a high ammonia abundance ($\sim4$x Solar N value) that are transported upwards in Jupiter's atmosphere to the cloud condensation layers. They interpreted these plumes to be the upwelling counterpart to the 5-micron hotspots formed in an equatorial Rossby wave, according to theories of \citet{2000_Showman} and \citet{2005_Friedson}. 3) They showed that Jupiter is extremely dynamic at pressures $<2-3$ bars, where numerous vortices, including the Great Red Spot (GRS) and Oval BA, are visible. Many of these were seen both at radio and in optical images. 

In this paper, we examine separate observations in a more restricted frequency range, with the specific goal of determining the properties of turbulence, atmospheric waves, and connection of radio features' structure to those seen at optical wavelengths. We particularly explore the dynamics of Jupiter's equatorial region where our resolution is the best and wind speeds are sufficiently high to allow features to translate measurable distances over short time spans.

Section 2 details our VLA observations and the data reduction techniques used to produce a cylindrical map of Jupiter at radio wavelengths. In Section 3 we describe a search for waves and review Kolmogorov theory. In the following subsections we present analysis of derived power spectra and power spectra fits for various latitudes and compare our techniques and results to similar studies at different wavelengths. We focus on the (GRS) region in section 4 and compare the radio map to an optical mosaic spanning the same area. Section 5 details analysis where we compare the radio flux density meridional profile to zonal winds and optical intensity profiles. Section 6 focuses on the dynamics at optical and radio wavelengths of Jupiter's northern equatorial region. Particularly, we investigate the connection of radio hotspots to the equatorial wave associated with the 5-micron hotspots and regions near the NEB. 

\section{Observations}

We observed Jupiter at wavelengths near 2 cm with the VLA in four separate sessions on February 3 (02:50-04:50 UTC and 08:20-12:50 UTC) and 4 (02:45-04:15 UTC and 11:20-12:50 UTC), 2015. The four sessions were arranged such that we achieved complete longitudinal coverage of the planet as it rotated, to facilitate construction of a final cylindrical map. The spatial resolution of an interferometer like the VLA is determined by the observing frequency and the maximum spacing between its antennas. We observed with the VLA's Ku-band receiving system, covering the frequency range from $\sim$12 to $\sim$18 GHz (wavelengths from $\sim$1.7 to $\sim$2.5 cm). At the time of our observations, the VLA had nearly completed its move from the hybrid CnB-configuration to the full B-configuration, resulting in a longest spacing of $\sim$11 km in both the N-S and E-W directions. This translates to a spatial resolution of $\sim$0.42$^"$ at our central observing frequency, and a linear size at the distance of Jupiter ($\sim$4.35 AU) at the time of $\sim$1500 km ($\sim$1.5$^\circ$ at the sub-Earth point). In a similar vein, the largest angular scale to which an interferometer is sensitive is determined by its shortest baseline. Our shortest baseline was $\sim$200 m, corresponding to a largest angular scale of $\sim$12$^"$. At the time of observation, Jupiter was near opposition, and over 45$^"$ in diameter - much larger than this scale. Because of this we cannot determine spatial structure on scales larger than about 1/4$^{\rm th}$ of Jupiter's diameter and are also not able to retrieve absolute brightness temperatures from our data. Fortunately we are mostly concerned with the spatial structure at small scales for this investigation, and are not seriously hampered by this lack of sensitivity to large-scale structure.

We used a single observation of the calibrator 3C 286 (J1331+3030) in our second observing session to determine the absolute flux density scale, and for delay and bandpass calibrations. The absolute flux density scale is good to $\sim$3\% at these wavelengths \citep{2013_Perley}. We used the bright, point-like radio calibration source J0854+2006 to calibrate the time-variable complex gain of the antennas and electronics and to transfer the flux density calibration to the three sessions where 3C 286 was not observed. Our derived flux density for this source was 6.53 Jy at a frequency of 17.064 GHz, with a spectral index of 0.22. We assumed that it did not change significantly in flux density over the roughly 34 hour duration of observing, and simply assigned that flux density to the three sessions in which 3C 286 was not observed.  Because of this non-standard method of observing, we could only use the standard VLA calibration pipeline (https://science.nrao.edu/facilities/vla/data-processing/pipeline) for the session in which 3C 286 was observed. For the other sessions, we had to modify the pipeline to not expect 3C 286 to be observed, and to specifically set the flux density of J0854+2006, rather than deriving it by bootstrapping from 3C 286. Otherwise, the initial calibration was as for all other continuum VLA observations.

At the end of this initial calibration, we had data organized into 48 sub-bands, each of 128 MHz width, covering the range from $\sim12.0-18.0$ GHz, in full polarization. We needed to flag a small amount of RFI that remained after the default flagging. We then searched for background sources, such as the Galilean moons, by taking out the tracking of Jupiter in the visibilities and making a large image, but found none. At this point we combined the data into a longitudinally-smeared image of Jupiter, using a limb-darkened disk as the initial model \citep{1999_Butler}. This image was then used to self-calibrate the data \citep{1999_Cornwell}. We finally used the method of \citet{2004_Sault} to make a non-smeared or longitudinally resolved map of the planet, in Jovigraphic coordinates (used throughout the rest of this paper). The result is shown in Figure \ref{fig1}, where radio brightness temperatures are on a color scale with red as brighter (hotter) and blue as dimmer (cooler). The contrast was enhanced in the map by first subtracting a uniform limb-darkened disk (see e.g. de Pater 2016); the color scale in Figure \ref{fig1} has been offset by adding the brightness temperature of the disk that was originally subtracted - 168.76 K. Our final map spans the full $0^\circ-360^\circ$ in longitude and latitudes from 74$^\circ$S to 74$^\circ$N - higher latitudes are excluded as projection effects make the resolution and signal to noise (SNR) much poorer in those regions. The large scale intensity variations, most noticeable at high latitudes with a wavenumber of $k\sim3$, were artifacts generated due to the lack of short spacings sampled in the VLA's B-configuration.

Given the observation's intrinsic spatial resolution, we expect the linear resolution to be $\sim$1.5$^\circ$ at Jupiter's equator in the final map. A map of effective resolution size as a function of location on the planet is a by-product of the map production scheme of \citet{2004_Sault}, and this is shown in Figure \ref{figr}. The best resolution is $\sim$1.3$^\circ$ at the equator, decreasing to about $\sim$1.8$^\circ$ near $\pm60^\circ$ and worse still at $\pm70^\circ$ because of projection effects elongate the resolution element to be nearly $\sim$3.5$^\circ$.

As noted above, our wavelengths are mostly sensitive to ammonia abundance, so higher temperature areas (red in Figure \ref{fig1}) are generally regions of less ammonia abundance, where we see to higher pressures and hence warmer layers in the atmosphere. Concomitantly, lower temperature areas (blue in Figure \ref{fig1}) are regions with greater ammonia abundance, where we see high altitude cooler layers in the atmosphere. The weighting functions at the two extremes of our frequency coverage are shown in Figure \ref{figw} as a function of pressure. The majority of the emission we detected originates from a pressure range of $\sim0.5-2.0$ bars, with a peak around 1.1 bars.

\section{Wave Power Spectra Analysis}

\begin{sidewaysfigure}
\begin{center}
\includegraphics[scale=0.5]{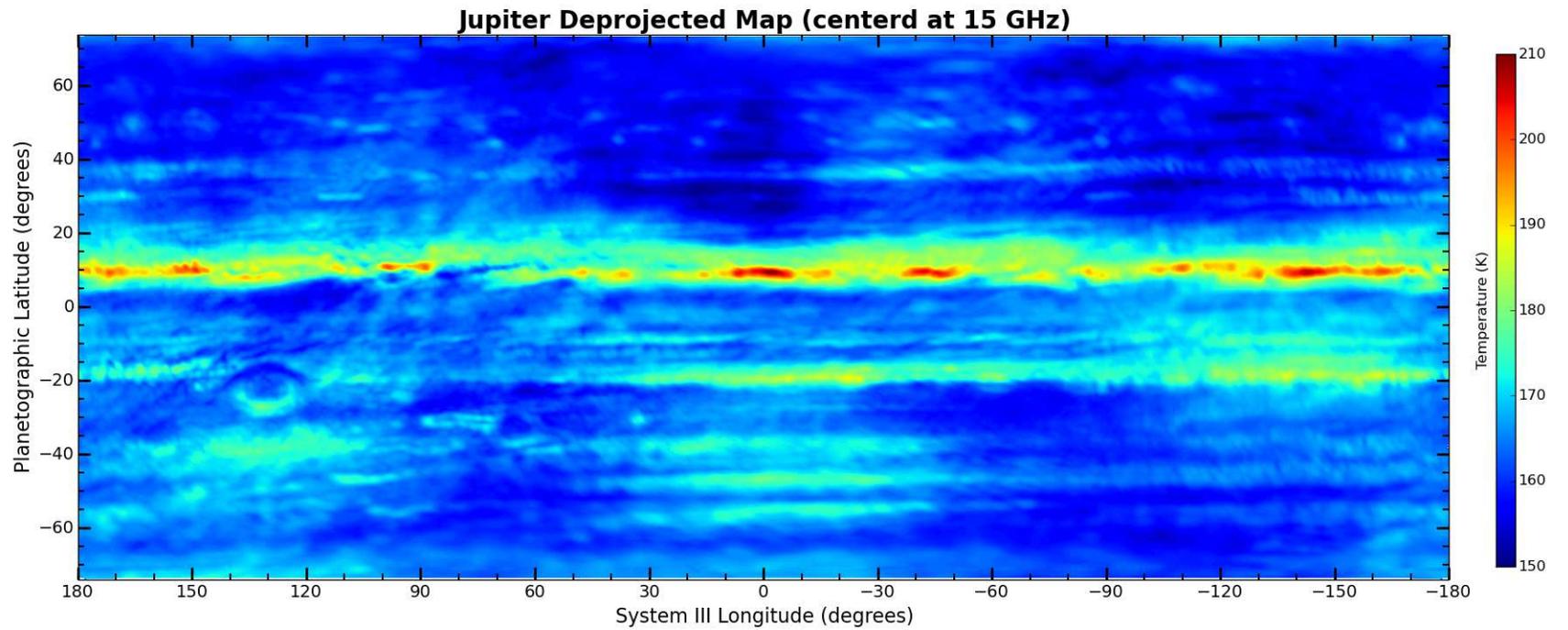}
\caption{Radio brightness map in Jovigraphic coordinates. These are deviations from a disk model, as described in the text. The color bar on the right indicates radio brightness temperatures - red for higher temperatures (low ammonia abundance) and blue for lower temperatures (high ammonia abundance). The $k\sim3$ features at high latitudes are artifacts produced from a lack of short spacings in the VLA's B-configuration. \label{fig1}}
\end{center}
\end{sidewaysfigure}

\begin{figure}
\begin{center}
\includegraphics[scale=0.5]{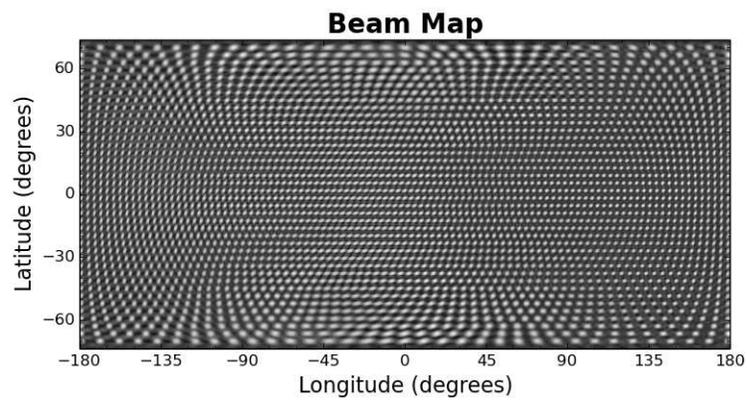}
\caption{Map of the effective resolution in Figure \ref{fig1} as a function of position on the planet. Intensity patterns represent many synthesized beam solutions for visibilities that were assembled together to generate this map. \label{figr}}
\end{center}
\end{figure}

\begin{figure}
\begin{center}
\includegraphics[scale=0.7]{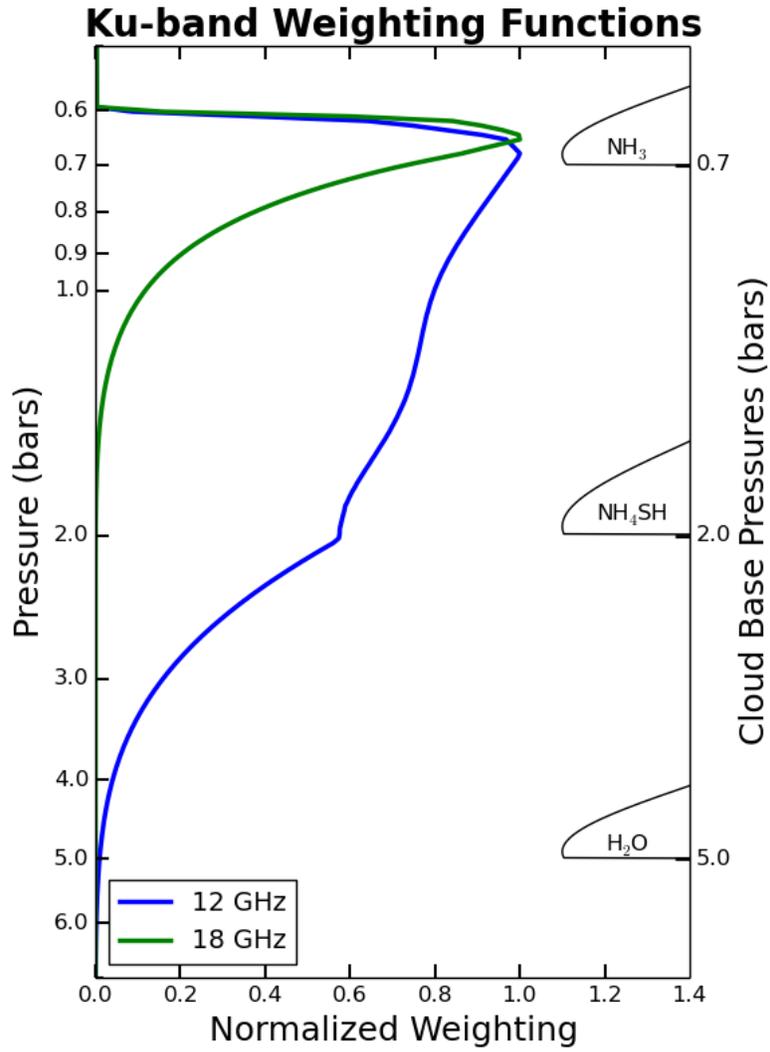}
\caption{Weighting functions at 12 and 18 GHz versus pressure in Jupiter's atmosphere based on ammonia profile 'a' in Fig. 3A of \citep{2016_dePater}. Cloud bases are also located at their predicted pressures according to condensation temperatures. \label{figw}}
\end{center}
\end{figure}

The most prominent feature in Figure \ref{fig1} is the bright radio-hot belt near 10$^\circ$N which dominates the spatial structure near the equator. \citet{2016_dePater} also showed that this radio-hot belt, at the interface of the NEB and NZ, was the most prominent feature at wavelengths up to $\sim6$ cm. The radio hotspots seen in Figure \ref{fig1} are areas of relatively lower ammonia abundance and are likely associated with the planetary wave seen in infrared observations at 4.9 microns and at optical wavelengths as dark areas between high altitude plumes \citep{1996_Orton, 2001_SimonMiller}. These radio hotspots and their potential connection to the infrared hotspots will be discussed in detail later but served as motivation to look for waves at all latitudes in Figure \ref{fig1}.

We searched Figure \ref{fig1} for wave structure using Fast Fourier Transform (FFT) analysis methods to calculate the complex response function at all latitude grid points over the entire 360$^\circ$ longitude span. We summed the complex response with line by line scans over a latitude range, here $\pm60^\circ$ degrees for the entire map. The summed complex response function was converted into FFT spectral power density and represents the longitudinal periodicity of 2 cm emission over Jupiter's 360$^\circ$ circumference for a particular latitude range. The spatial frequency or wavenumber ``$k$", denotes the number of complete sinusoids that best fit in the planet's circumference at a particular latitude. 

\begin{figure}
\begin{center}
\includegraphics[scale=0.8]{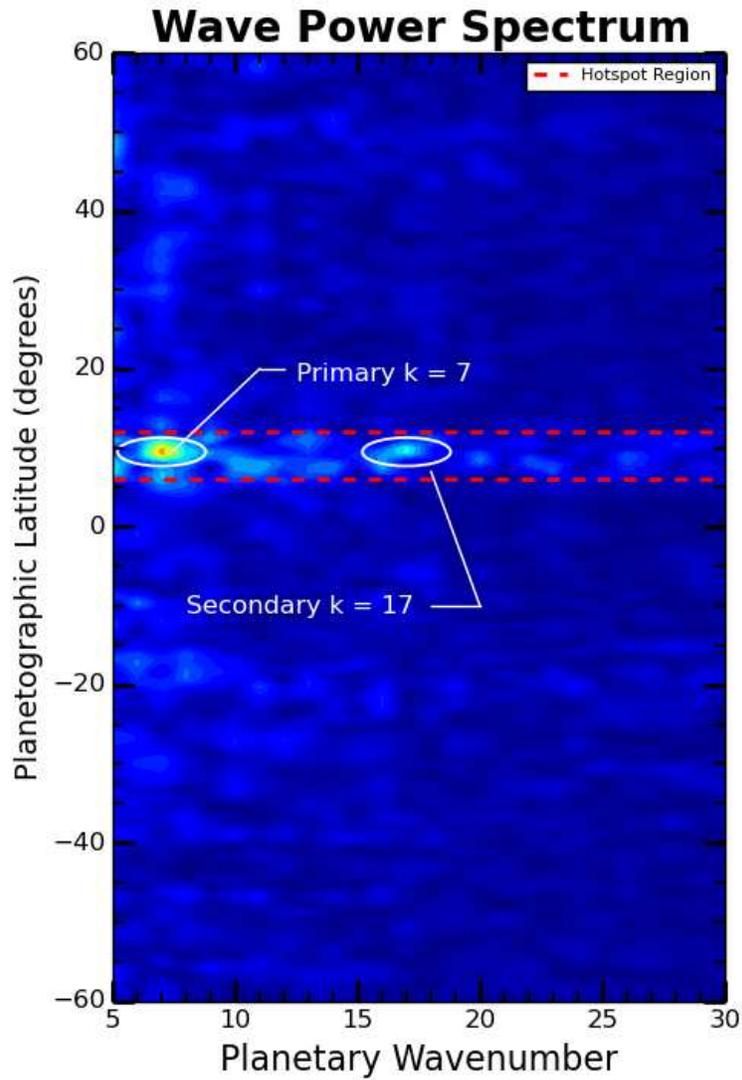}
\caption{FFT power plotted as latitude versus wavenumber in log power color scaling where red is highest in power, cycling through yellow, green and blue which is the lowest power. The primary and secondary wavenumber components of the hotspots wave feature near 10$^\circ$ are circled and labeled. The red dashed lines mark the latitude bounds used in further analysis of the map. \label{fig2D_wave}}
\end{center}
\end{figure}

The FFT power is shown as a surface map for latitude versus wavenumber in Figure \ref{fig2D_wave} where the color scale spans power in a log scale from low to high, blue to red. High wave power (red) indicates strong periodicity of a radio intensity signal at a given spatial frequency (wavenumber $k$), while low wave power (blue) indicates no periodicity in the signal. A high power peak is found at a primary wavenumber of $k = 7$, which is similar to the 5-micron hotspots' wavenumber presented elsewhere \citep{2000_Showman, 2005_Friedson, 2005_Vasavada}. Another peak in power at 10$^\circ$N appears at wavenumber $k = 17$, which could be attributed to the radio hotspots splitting \citep{2013_Choi}. We only display wavenumbers in Figure \ref{fig2D_wave} in the wavenumber range of $k = 5-30$. The low wavenumber limit is set by the largest angular size the VLA can resolve for our observation, as discussed earlier. The observation's resolution yields a high wavenumber limit determined to be $k = 114$. We cut off high wavenumbers in Figure \ref{fig2D_wave} greater than $k = 30$ because structure in the FFT power is greatly diminished for all latitudes.

\begin{figure}
\begin{center}
\includegraphics[scale=0.7]{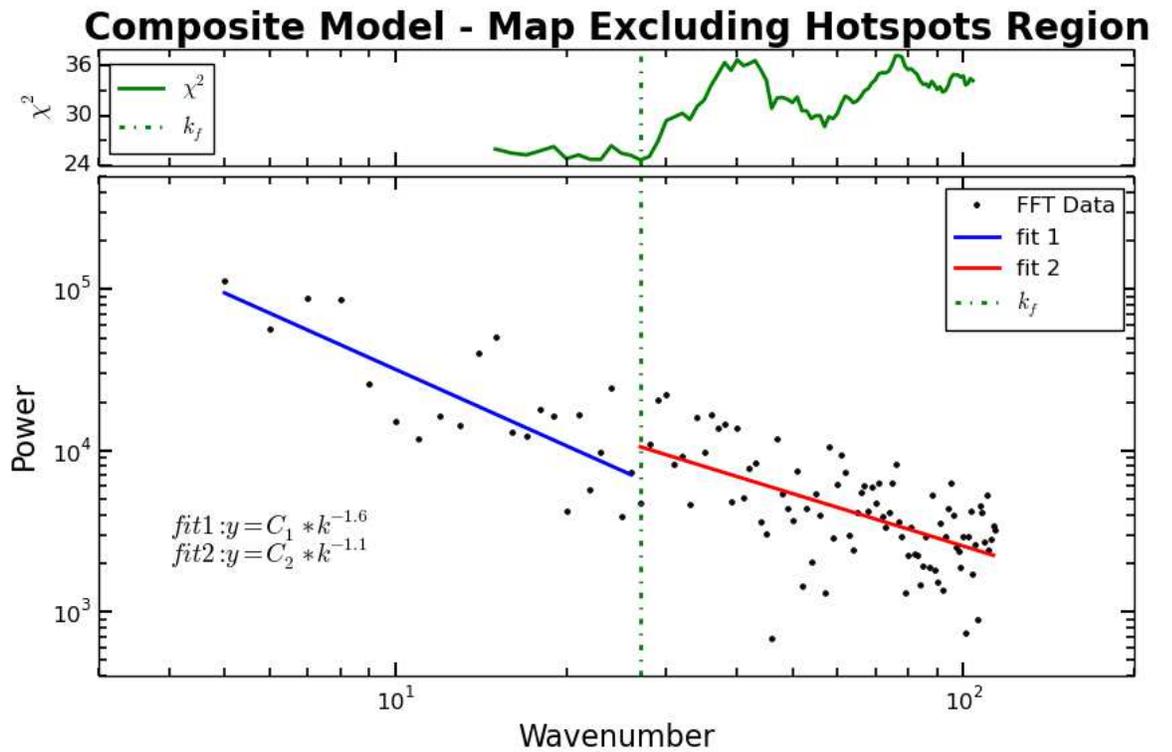}
\caption{Sample plot for fits to the power spectra of the map excluding the hotspots region latitudes. Fits 1 and 2 are shown in blue and red (respectively) over their fitted wavenumber ranges. The green curve displays $\chi^{2}$ and the vertical dashed line represents the forcing scale wavenumber $k_f$ that minimized the best composite model fit to the power spectra. \label{figspectra-fit}}
\end{center}
\end{figure}

Figure \ref{fig2D_wave} depicts wave information over all latitudes and wavenumbers simultaneously but one is able to further investigate more detailed properties of a spectrum if it is displayed for a specific latitude, or range of latitudes, in a plot of power versus wavenumber in log-log space. We wish to determine spectral indices (or slopes), changes in spectral indices and identify kinks (or breaks) in spectra. A sample spectrum showing power versus wavenumber for the entire map, excluding the hotspots region labelled in Figure \ref{fig2D_wave}, is shown in Figure \ref{figspectra-fit}. 

Past studies of passive tracer spectra at optical and infrared wavelengths have found transition wavenumbers in the range of $k_f = 80-300$, with associated uncertainties of $\delta k_f = \pm20$ \citep{1996_Harrington, 2009_Barrado, 2011_Choi}. A break in a spectrum is an indication of atmospheric forcing at that particular wavenumber and is commonly referred to as the forcing scale ``$k_f$". The previous studies report a large range for the forcing scale $k_f$ in Jupiter's atmosphere leaving its value highly unconstrained at different pressures in Jupiter's atmosphere. We decided to investigate the wave power spectral properties of Figure \ref{fig1} at various latitudes according to latitude boundaries that correspond to eastward jets, westward jets, and the radio hotspots region. We compare the results to a previous infrared study \citep{1996_Harrington} and to higher resolution optical studies \citep{2009_Barrado, 2011_Choi}.

\subsection{Predicted Power Spectra}

In order to study the dynamics of Jupiter's atmosphere at a particular pressure range, one can take a map of the planet at a wavelength that probes that pressure and construct a wave power spectrum of FFT energy density versus spatial scale, wavenumber $k$, for the map. Our VLA observation senses ammonia abundance distributions, which can be utilized to conduct analysis of a passive tracer power spectrum. This is separate from the kinetic energy power spectrum, which is formed by conducting similar FFT analyses on the sum of the squared zonal and meridional winds versus wavenumber \citep{2006_Vallis}. The properties of a spectrum give an indication of the turbulent dimensionality of the atmosphere at the probed pressure range. If the winds beneath the weather layer are connected to the winds at the cloud layer (implying connected winds as suggested by Sanchez-Lavega et. al. 2008), one might expect the turbulence to be 3-D. Conversely, if those winds are not connected to those at the cloud layer, one might expect the turbulence to be 2-D. Determination of this dimensionality thus helps distinguish between these two models for the dynamics of the weather layer on Jupiter.

\begin{figure}
\begin{center}
\includegraphics[scale=0.6]{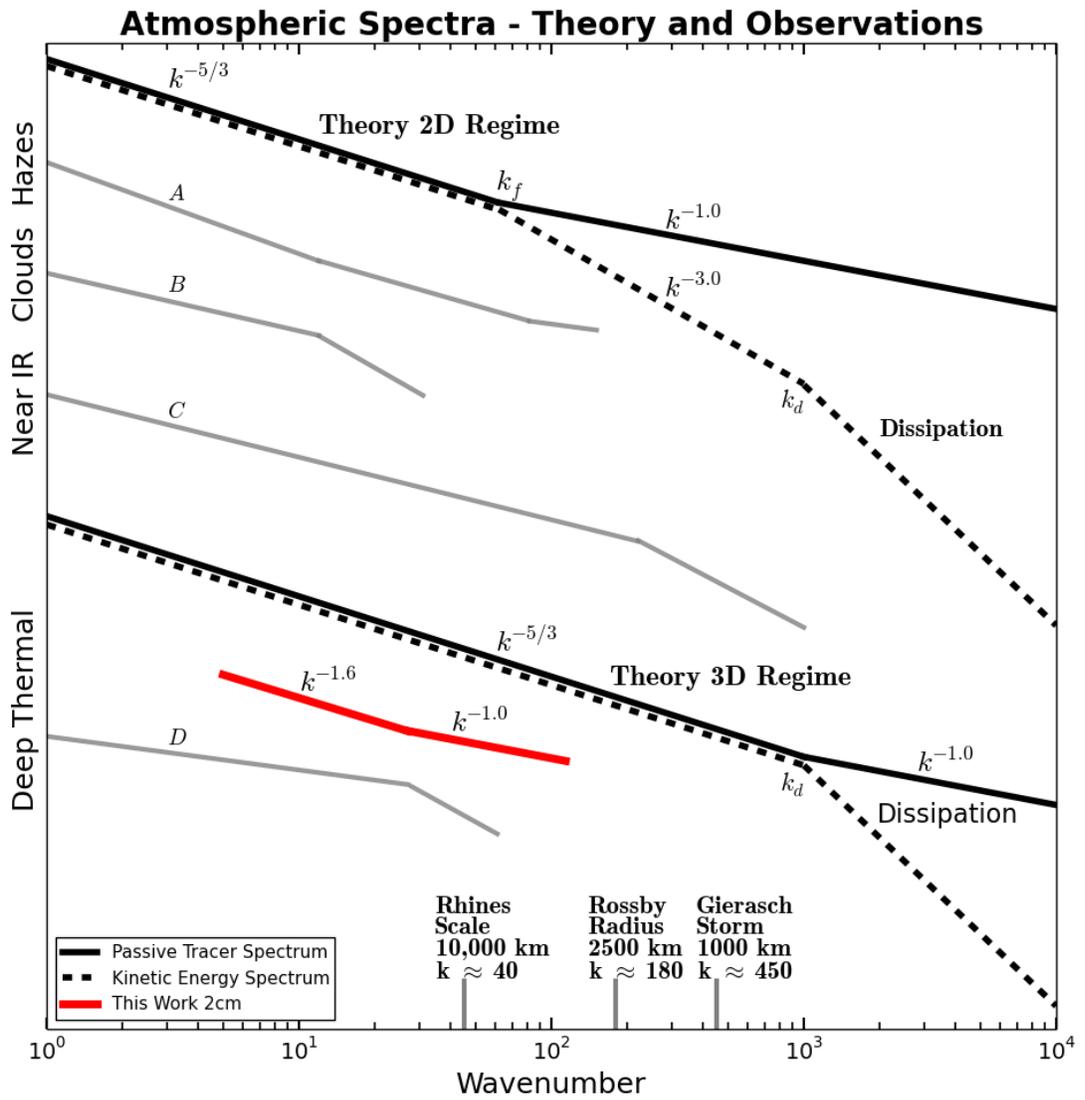}
\caption{Log-log space plot of sepctral power (shown also in pseudo atmospheric depth) versus wavelength for various Jovian observations of passive tracer spectra. The presence of a kink in most spectra occur somewhere in the predicted length scales identified. Selected power spectra from previous studies are lettered along with their corresponding observational wavelengths: A - \citet{2008_Barrado} 258 nm, B - \citet{1985_MitchellMaxworthy} 420 nm, C - \citet{2011_Choi} 756 nm, D - \citet{1996_Harrington} 4.9 $\mu$m. Our derived power spectra is shown in red. \label{obs_spectra-theory}}
\end{center}
\end{figure}

Kolmogorov theory shows that an atmosphere in a 3-D turbulent regime should have the passive tracer and kinetic energy spectra that have spectral slopes of $k^{-5/3}$ for increasing wavenumbers $k$. This is caused by the transfer of energy from large to small scales or low to high wavenumbers \citep{1986_Nastrom, 2006_Vallis}. A single change in the slope will occur at a characteristic wavenumber $k_d$, where small scale diffusion dominates turbulence; for wavenumbers greater than $k_d$, the passive tracer power spectrum will have a slope following a relationship of $k^{-1}$. The kinetic energy power spectrum differs only for wavenumbers greater than $k_d$, where it follows a relationship of $k^{-n}$ with $n > 3$ in the energy dissipation regime. Figure \ref{obs_spectra-theory} shows the theoretical passive tracer and kinetic energy power spectra with corresponding labels in the lower portion of the plot for a 3-D turbulent atmosphere.

In a 2-D turbulent regime with forcing at some characteristic wavenumber $k_f$, there is a similar $k^{-5/3}$ relation as for the 3-D regime for wavenumbers less than $k_f$. This is caused by energy being injected into the atmosphere at the forcing wavenumber and being transferred from higher to lower wavenumbers (inverse energy cascade). In the passive tracer case, for wavenumbers greater than $k_f$, the power spectrum follows a $k^{-1}$ relation. In the kinetic energy case, the power spectrum follows a $k^{-3}$ relation between $k_f$ and $k_d$, then experiencing significant energy dissipation, and as before follows $k^{-n}$ with $n > 3$ for higher wavenumbers. Figure \ref{obs_spectra-theory} shows the theoretical power spectra for a 2-D turbulent atmosphere with its corresponding labels near the top of the plot.

\subsection{Infrared Power Spectra Comparison}

We can compare our observed passive tracer power spectrum produced from Figure \ref{fig1} to thermal emission power spectra presented in \citet{1996_Harrington}. We fitted linear power laws to our power spectra in log space over the same wavenumber ranges as in \citet{1996_Harrington}. They investigated two wavenumber ranges, low $k = 1-24$ and high $k = 28-60$, for different latitude regions according to the observed zonal wind profile described by \citet{1986_Limaye}. \citet{1996_Harrington} investigated the power spectra of the Jovian eastward and westward jets separately but also analyzed their entire map's power spectrum. They identified, via visual inspection, a kink in their infrared power spectra at wavenumber $k_f\sim$26, which served as the transition between the two wavenumber ranges. Although we do not identify a similar kink, we analyzed the same wavenumber and latitude ranges to compare radio power spectra to the analysis presented in \citet{1996_Harrington}. 

To investigate the spatial wave structure, \citet{1996_Harrington} performed Lomb-Normalization on their data, because it contained gaps in spatial coverage, to construct power spectra from the generated periodograms. FFTs do not accurately determine power for data sets that contain gaps or are unevenly spaced, and will artificially overestimate power at higher frequencies to fit sharp discontinuities in an effect known as ``ringing". Our cylindrical map in Figure \ref{fig1} has evenly spaced data so we can utilize FFTs to generate our power spectra, as described previously. \citet{1996_Harrington} also averaged over $1^\circ$ degree latitude strips to increase SNR, which is also something we did not need to do given the VLA's sensitivity and Jupiter's strong emission at 2 cm.

We compared the entire map's power spectra slopes for a limited latitude range $\pm 60^\circ$ and also the slopes for latitude ranges specified in \citet{1996_Harrington} for eastward and westward jets. We additionally modified the latitude transitions of the zonal wind profile according to the Outer Planet Atmospheres Legacy program (OPAL) HST observation taken on January 19, 2015, about two weeks prior to our observation \citep{2015_Simon}, and found little variation in our derived spectral slopes. The results of our analysis are compared and presented in Table \ref{table1} with uncertainties estimated with bootstrapping techniques.

The spectral slopes found by \citet{1996_Harrington} over the low wavenumber range, $k=1-24$, were -0.7 for both eastward and westward jets while we find steeper slopes near -1.0 over a wavenumber range of $k=5-24$. In the high wavenumber regime, our slopes are consistently shallower than those found in \citet{1996_Harrington} and when we extend the fits to our high wavenumber limit, the slopes became even shallower. We do find a difference in slopes between the two wavenumber ranges, similar to \citet{1996_Harrington} but not nearly as significant. We also find a clear distinction, given our uncertainties, in the jet's spectral slopes for low and high wavenumber ranges found by \citet{1996_Harrington}, implying zonal jets produce unique spatial power spectra when observed at 4.9 microns and 2 cm.

A few caveats of our comparison to the infrared study by \citet{1996_Harrington} must be noted. First, we are comparing wave power spectra created by two different analysis techniques - \citet{1996_Harrington} used Lomb-Normalization while we utilized FFTs. We created synthetic power spectra, both with and without gaps in the data, and linear fits using Lomb-Scargle analysis consistently did not retrieve the true synthetic spectral slopes while FFT analysis did. Secondly, the infrared and radio wavelengths observational pressure ranges are sufficiently different that the wave power spectra are not sampling similar vertical layers in Jupiter's atmosphere. We also do not observe a kink at the same location in the power spectrum presented in \citet{1996_Harrington} and this prompted us to explore other power spectra fitting techniques as discussed in the following section. 
  
\subsection{Optical Power Spectra Comparison}

Studies by \citet{2009_Barrado} and \citet{2011_Choi} investigated power spectra derived from maps produced by optical wavelength HST and Cassini observations. These observations are limited to pressures lower than the thermal emission we observed, mainly exploring the troposphere near the ammonia-ice cloud upper deck. They analyzed their power spectra with a composite model of two linear fits in log space, where the transition between the two fits was a free parameter designated as $k_f$. The low wavenumber and high wavenumber linear fits have power law slopes of $m_1$ and $m_2$, respectively. The transition between $m_1$ and $m_2$ occurs at $k_f$, which is determined over the entire wavenumber range such that $k_f$ minimizes $\chi^2$ of the composite model fits to the power spectra data. We applied this technique to our radio power spectra and we found that most composite model fits, over the wavenumber range $k = 5-114$, determined $k_f \approx13-36$. All results from our composite model analysis for different power spectra are presented in Table \ref{table2}. 

The composite model spectrum fit for the map excluding the hotspots region, latitudes ranging $8-12^\circ$N is shown in Figure \ref{figspectra-fit}. The blue and red lines are the low and high wavenumber fits, respectively, with the transition wavenumber $k_f = 27$ identified by the green vertical center line. This passive tracer power spectrum reveals that Jupiter's atmosphere has spectral slopes that best match the characteristics of a 2-D turbulent atmosphere. 

Presented in Table \ref{table2} are estimated uncertainties in spectral slopes and forcing scale wavenumber $k_f$ via bootstrapping techniques. We also varied the minimum model segment lengths and report the converged wavenumber array lengths, $k_{mod}$, used in fitting the power spectra in Table \ref{table2}. Additionally, we also compared the output of the fitting algorithm from \citet{2009_Barrado} and \citet{2011_Choi} to a routine we developed. Our routine identifies the transition between the two linear fits at the intersection of the fitted slopes and then calculates $\chi^2$ for this new composite model fit to the power spectra. The best fit model is then selected from the minimized $\chi^2$ for this intersection calculated forcing scale. We found this intersection fitting algorithm produced forcing scales and slopes that were consistent (within the bootstrapping uncertainties) with the earlier results from the composite model power spectra fitting techniques from \citet{2009_Barrado} and \citet{2011_Choi}.

\section{Nearly Concurrent Radio and Optical Observations}

\begin{sidewaysfigure}
\begin{center}
\includegraphics[scale=0.4]{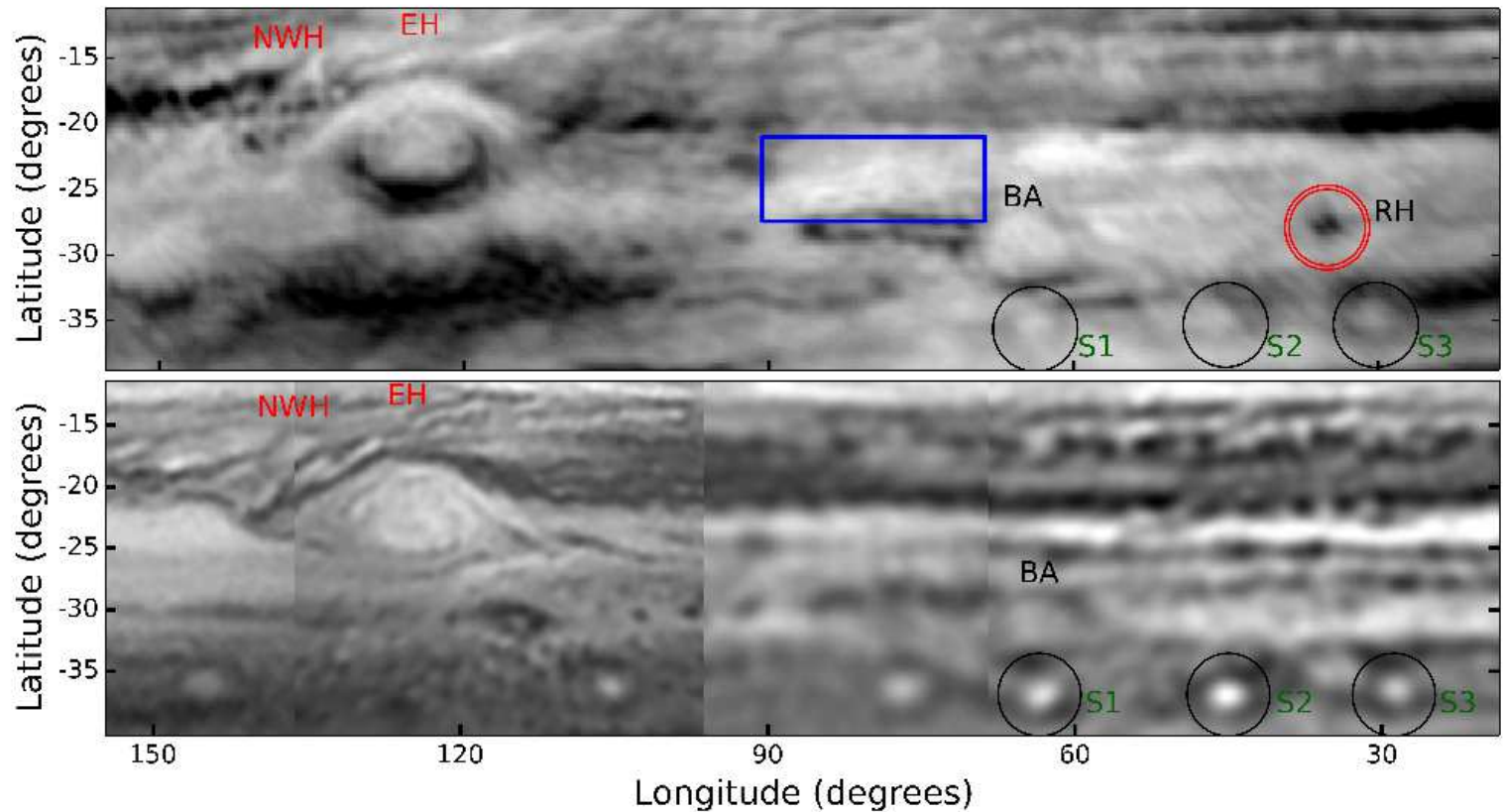}
\caption{Upper plot is the radio map (inverted and contrast enhanced) and below is an optical wavelength mosaic from ALPO library near the same times as the VLA observation. Spots S1, S2 and S3 are circled in black while the radio hole RH is circled in red twice. The GRS on left west portion of plot has its wake features, Northwest Hallow (NWH) and Equatorial Hallow (EH), labeled in both panels northward to their respective features. Oval BA is labeled north of the storm in both panels while in the top panel a blue rectangle outlines a region of recent upwelling. \label{fig-vis-rad}}
\end{center}
\end{sidewaysfigure}

We investigated features seen in Figure \ref{fig1} in the radio to nearly concurrent optical wavelength observations. The lower panel of Figure \ref{fig-vis-rad} shows an optical wavelength mosaic produced from three different amateur images taken on February 1, 2015 (Anthony Wesley, Australia), February 2, 2015 (Torsten Hansen, Germany), and February 4, 2015 (Michel Jacquesson, France) and posted in the Association of Lunar and Planetary Astronomers (ALPO) website. The upper panel in Figure \ref{fig-vis-rad} is an inverted radio map, where dark areas are low opacity ammonia regions, corresponding to emission from the 1-2 bar pressure level in the atmosphere - higher temperatures in Figure \ref{fig1}. Conversely, light areas are high opacity ammonia areas near the upper cloud layer of Jupiter's atmosphere - lower temperatures in Figure \ref{fig1}. This radio representation in gray scale more closely resembles the optical mosaic where light areas are reflective clouds with high albedo that are presumed to be at greater altitudes than darker areas with low albedo at lower altitudes.

The most notable feature in Figure \ref{fig-vis-rad} is the Great Red Spot (GRS). The GRS has a interesting meridional asymmetry seen at radio wavelengths, present infrared wavelength observations (notably 5 microns), which indicates there is a higher abundance of ammonia on its northern flank \citep{2010_dePater, 2010_Fletcher, 2016_Fletcher}. This asymmetry supports prior optical wavelength observations that the GRS is a tilted ``pancake" \citep{2002_SimonMiller} and further suggests that its tilting extends to our observation's pressure range.  

The GRS is bounded on its northern edge by a westward jet and an eastward jet on its southern edge and specific features related to the dynamics in GRS region are labelled in both panels of Figure \ref{fig-vis-rad}. Without observations to track the motion of features in consecutive radio maps, we assume the zonal wind profile obtained at optical wavelengths is the same at the pressure range of our observation. Adjacent to the GRS is a large wake whose prominent structure is slightly different in the two panels. In the optical, there is more extended structure around the GRS, which is less obvious at radio wavelengths. However, the features known as the Northwest Hollow (NWH) and Equatorial Hollow (EH) are seen in the radio panel on the north-west and northern edges of the GRS, consistent with past observations at optical wavelengths and modeling studies \citep{2013_Morales}. These features may be the result of the jet's deflection around the GRS or the motion of the GRS as it moves westward; either case it is clear that the GRS impacts the flow on its northern edge to at least the high end pressure range of our radio observation.

The oval BA anti-cyclone is labelled to its immediate north in both panels of Figure \ref{fig-vis-rad}. A blue rectangle outlines an area Northwest of oval BA identified as a site of abundant ammonia and an indication of recent local convection. This is the same type of region that New Horizons detected transient spectrally identifiable ammonia clouds (SIAC) as signs of up-welling on the outer edges of a cyclonic cell \citep{2007_Reuter}.

Smaller spots, labelled $S1-S3$, are circled in Figure \ref{fig-vis-rad} and show good correlation indicating that the vertical structure of each spot spans altitudes probed by both wavelengths. Closer inspection of Figure \ref{fig1} shows these spots as cool regions implying high levels of ammonia. Interestingly, there are 3 other smaller spots seen in the optical mosaic that show no clear radio counterpart. These particular storms are shallow and their lower extent likely constrained to the optical cloud deck layer near 0.7 bars. At this latitude, storms at different longitudes appear to span different pressure ranges. It is possible that the GRS impacts the southern route of the spots and forces them upward to higher altitude leaving no radio emission signature and after passing the GRS's longitude, descend downward in altitude again.

Not all features observed in the radio map have clear optical counterparts. One such radio feature is identified as a radio hole (RH), circled twice in red and labelled in Figure \ref{fig-vis-rad}. The RH has a diameter of roughly $3^\circ$ and is seen as a warm feature in Figure \ref{fig1}. The RH could be produced by local convection or sinking air that is decoupled from the dynamics of the atmosphere above it. Regardless of the clearing mechanism responsible for the creation of the RH, its presence indicates that features at the pressures probed by the radio observation are influenced by dynamics that are not necessarily coupled to the clouds observed at optical wavelengths.  

\section{Meridional Radio Structure, Zonal Winds and Banding}

Besides the GRS, one of Jupiter's most defining characteristics is its well known banded structure seen at optical wavelengths. A classical interpretation is that Jupiter's banding is produced by regions of rising and sinking air in the zones and belts, respectively \citep{1969_Ingersoll}. More recent studies have found evidence to support the classical view \citep{2015_Bjoraker, 2016_dePater}, while others suggest an alternative view of the large scale vertical motion to be reversed \citep{2000_Ingersoll, 2003_Porco, 2006_Showman}. As research on this topic progresses, more complicated treatments of Jupiter's large scale circulation \citep{2005_Showman} might be favored as we continue to compile new observations of moist convection and lightning \citep{1999_Little, 2000_Gierasch}.     

Our observation does not have the ability to track this large scale motion but does investigate the structure of the zones and belts relative to each other on Jupiter. Zones, at optical wavelengths, are high albedo bright regions with high altitude ammonia rich clouds. By contrast, belts are darker at optical wavelengths and are presumed to be regions of sparse cloud cover, revealing Jupiter at lower altitudes. Zones and belts are latitude bands in the zonal wind profile that correlate with its first derivative or regions of anti-cyclonic and cyclonic shear zones, respectively \citep{2005_Vasavada}. 

\begin{figure}
\begin{center}
\includegraphics[scale=0.5]{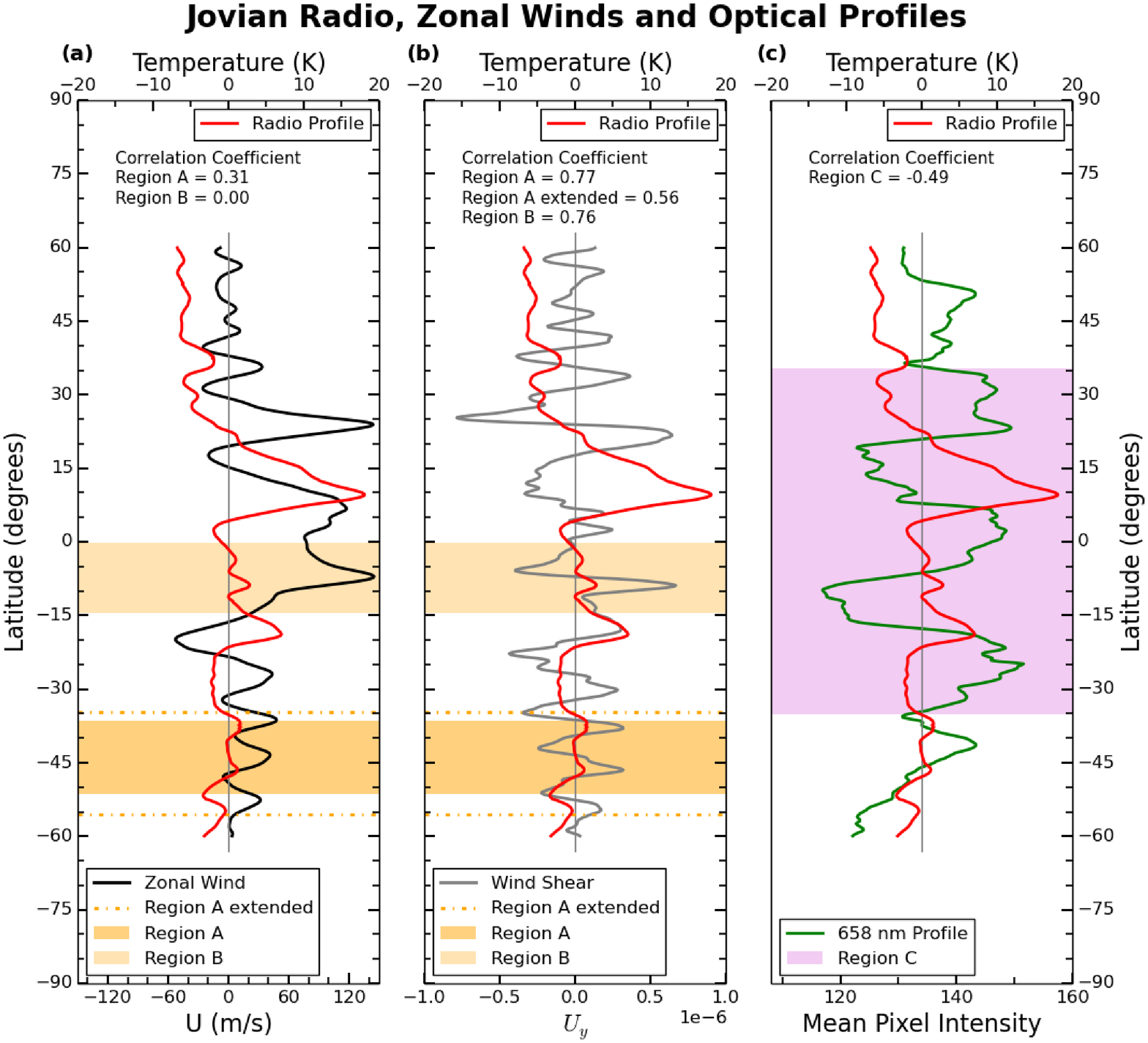}
\caption{The meridional radio profile from averaging emission across all longitudes in \ref{fig1} is shown in red in all three panels. Panels (a) and (b) show the zonal wind profile and its first derivative respectively, where shaded and outlined as described in the text. Panel (c) shows the averaged optical intensity profile for a specific HST filter wavelength where a shaded equatorially centered tropical region is identified. The corresponding correlation coefficients are displayed near the top of each panel for specific regions. \label{VLA_OPAL}}
\end{center}
\end{figure}

The radio intensity was summed and averaged over the entire longitude span in Figure \ref{fig1} and we compared this profile to the zonal winds and optical banding profiles from the OPAL 2015 observation \citep{2015_Simon}. The meridional radio profile is shown in panels (a-c) of Figure \ref{VLA_OPAL} in red, where the OPAL 2015 zonal wind profile is shown in black in panel (a) and its derivative in grey in panel (b). We did not find any strong correlations or anti-correlations over the entire latitude span when we compared these profiles, but upon closer inspection, we did see some correlation over certain latitude ranges.

We calculated the Pearson and Spearman correlation coefficients and associated p-values for each pair of variables for different regions and present our results in Table \ref{table3}. Panels (a) and (b) in Figure \ref{VLA_OPAL} show a highlighted southern region, ``Region A", that spans latitudes from roughly $52^\circ$S to $36^\circ$S. The zonal wind derivative, $U_y$, and the radio profile correlation was maximized over the latitude domain of Region A yielding a moderately strong coefficient of 0.77 and an associated p-value of 8e-14. We explored the sensitivity of the correlation analysis to the latitude domain and expanded Region A to include the three southern hemisphere zonal jets in the range from roughly $55^\circ$S to $35^\circ$S; identified by center lines north and south of Region A. We identify this region as ``Region A extended" and we find the correlation coefficient decreases moderately, but still agrees with a moderately strong correlation value of 0.56 and an associated p-value 8e-9. The extremely low p-values associated with the Pearson correlation coefficient represent the probability that the correlation is purely observed by two random datasets yielding the relationship for all pairs of data. The lower the p-value means a greater likelihood of a real and significant relationship between the two data sets, here the zonal wind shear $U_y$ and the radio profile intensity. 

``Region B" in panels (a) and (b) in Figure \ref{VLA_OPAL} is another latitude range, from roughly $15^\circ$S to the equator, where we found a strong correlation between the radio profile and $U_y$. The corresponding correlation coefficient 0.76 and p-value of 4e-13 suggest the relationship found in Region B is similar to Region A. 

It is interesting that this correlation exists in the southern equatorial zone, but is not present in the northern equatorial zone. It suggests that the zonal winds and known wave features seen in multiple wavelength images that reside at these latitudes have complicated wave-mean flow interactions that extend to pressures probed by the observed wavelengths \citep{2012_SimonMiller, 2013_Choi}. We will investigate the NZ, the NEB, and their connection to the 5-micron hotspots and radio hotspots features seen at these latitudes in the following section. 

We investigated the correlation of radio banding to optical banding, also from the OPAL 2015 observation, shown in panel (c) of Figure \ref{VLA_OPAL}. We performed correlation analysis on a latitude region, ``Region C", spanning $35^\circ$S to $35^\circ$N for the two intensity profiles. There is a moderately strong anti-correlation of -0.49 with a p-value of 7e-28, between the HST optical profile and the radio profile in this region where we revisit the zone and belt model. The lighter zones, centered at the equator and at $\pm25^\circ$, are anti-correlated with low radio flux and regions where ammonia is in high abundance. If high altitude ammonia rich air is present in these regions, then radio emission is impeded, and hence produce regions with low radio brightness temperatures. This is exactly what we find in our analysis when comparing zone latitudes in optical and radio wavelengths. The zone-belt model structure also explains the darker belts centered at $\pm15^\circ$ as low altitude cloud regions with radio bright regions. Since belts have few high altitude clouds, there is less ammonia gas to block thermal emission from higher pressures. 

The southern hemisphere radio banding we identified as Region A extended was also mentioned over a similar latitude range from $40^\circ$S to $60^\circ$S in \citet{2016_dePater}, where they found no correlation when comparing optical and radio meridional intensities. We also do not see this correlation, but we have shown that the zonal wind derivative $U_y$ is correlated to the radio banding over a similar latitude range. The southern hemisphere regions we identified do suggest the radio profile is more coupled to the dynamics observed at optical wavelengths than the northern hemisphere. The models to explain Jupiter's banded appearance, including the dynamics of belts and zones at optical wavelengths, are not apparently ubiquitous at the pressures probed by our radio wavelength observation; most notably, the lack of significant correlation in the northern hemisphere. 

\section{North Equatorial Zone/Belt Dynamics}

\begin{sidewaysfigure}
\begin{center}
\includegraphics[scale=0.45]{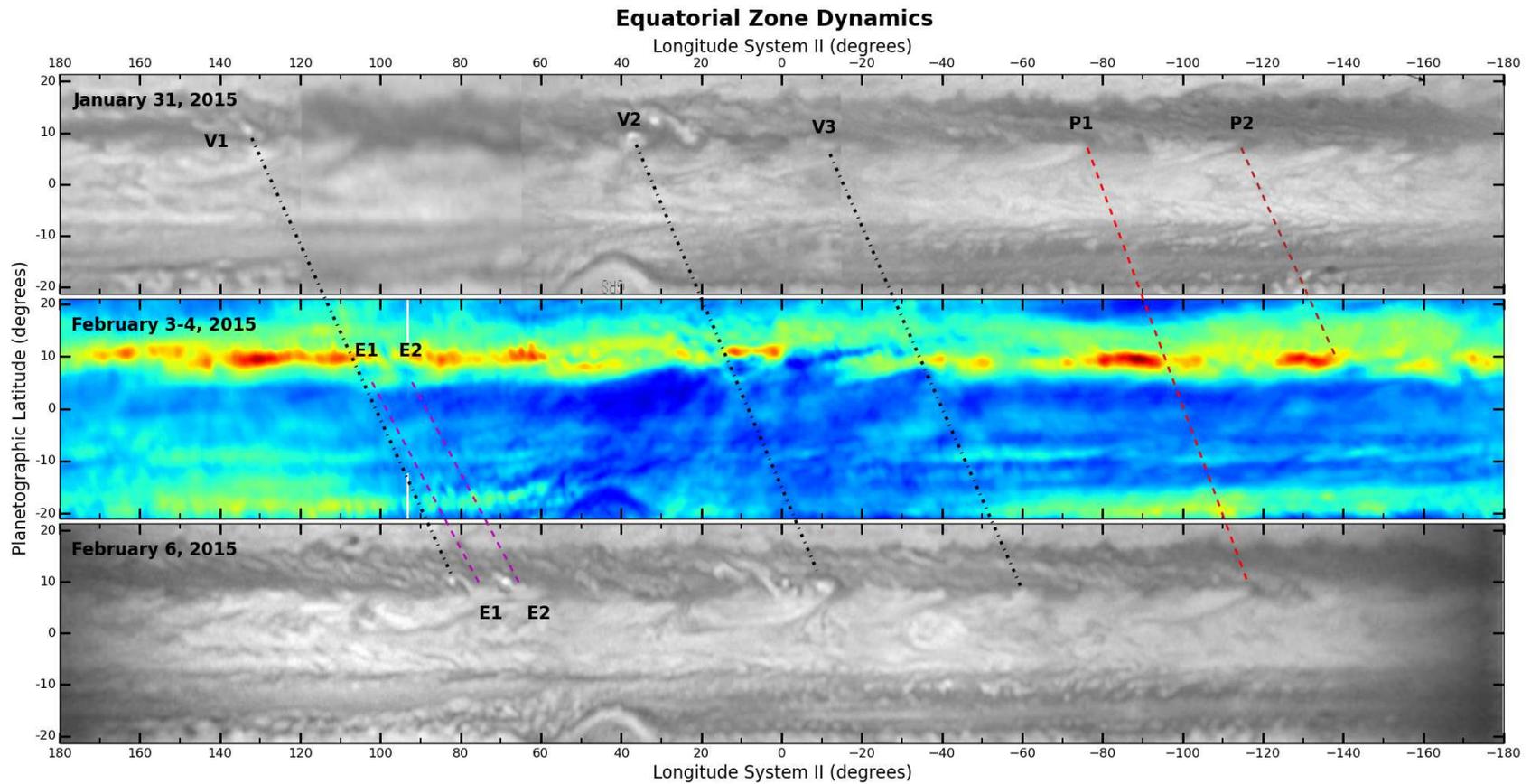}
\caption{The upper left corner of each panel is identified with is respective observation date. The top and bottom panels are optical maps constructed from amateur observations zoomed in towards the equatorial region. The middle panel is the radio observation, here shifted to System II longitude coordinates. Visible features $V1-V3$ are tracked in the optical panels and their velocity trajectories pass through or near radio counterparts in the middle panel. Plume $P1$ is also identified in all three panels while Plume $P2$ and equatorial features $E1$ and $E2$ are matched between two panels. The eastern edge longitudes and latitudes are (format - $Feature$[x1,y1]$\rightarrow$[x2,y2]): $V1$[131.5,11.0]$\rightarrow$[80.9,10.5]; $V2$[34.4,10.0]$\rightarrow$[-13.4,9.5]; $V3$[-13.2,8]$\rightarrow$[-63.9,8.0]; $E1$[100.6,7.0]$\rightarrow$[70.5,7.0]; $E2$[90.4,7.0]$\rightarrow$[59.9,7.0];\label{VLA_EQ}}
\end{center}
\end{sidewaysfigure}

We also investigated the dynamics of the EZ and NEB at optical and radio wavelengths. Without consecutive radio maps paired with simultaneous HST observations, we can not independently track moving features at both observation wavelengths to derive speeds. We also can not investigate vertical wind shear by analyzing simultaneous different HST filters \citep{1999_Simon} combined with radio wavelengths. However, we were able to find amateur optical observations of Jupiter made roughly three days prior to and following our radio observation. We created maps from amateur images taken on January 31, 2015 (Tiziano Olivetti, Italy) and February 6, 2015 (Hideo Einaga, Japan) and shifted the radio map into System II longitude coordinates (amateurs utilize System II where the GRS is the prime meridian) and focused our analysis on Jupiter's equatorial region from each observation as shown in Figure \ref{VLA_EQ}.

The top panel of Figure \ref{VLA_EQ} has three bright, white features labelled $V1-V3$, likely recent sites of convection, that move eastward from the top to bottom panel. The eastern most edges of $V1-V3$ were located in the top and bottom optical panels and the coordinates are connected via trajectory lines that span over the radio map in the center panel. The trajectories connecting features $V1-V3$ are nearly parallel lines and the features have calculated speeds in the range of $110-120\rm ms^{-1}$. The trajectories connecting features $V2$ and $V3$ intersect radio counterparts in the middle panel seen as blue circular features. Since the optical features are bright, white, ammonia rich clouds, they correspond to low temperature, high altitude features in the radio map (blue).

There are two equatorial features, identified as $E1$ and $E2$ in the middle radio panel as blue circular objects, that associate to a pair of optical features present in the bottom optical panel. We plot their trajectories between the middle (cool radio objects) and bottom (optically bright circular clouds) panels, excluding the top panel since the pair of features are not simultaneously identifiable, and calculate a speed of $100\rm ms^{-1}$. Their latitude is slightly more southward than features $V1-V3$ and they have a lower velocity, coincidently suggesting their motion is influenced by the equatorial Rossby wave that has long existed there. In general, it is difficult to disentangle the wave motion from the zonal jet speed at $7.8^\circ$N because correlation analysis at optical wavelengths tends to favor tracking the spatially large and coherent motion of the plumes and dark patches between them \citep{2013_Choi}. Individually tracking features at the pressures from consecutive 2 cm radio observations combined with optical could offer new insight into this region's dynamics \citep{2013_Choi}.   

The features labelled $P1$ and $P2$ are equatorial plumes, speculated to be connected to the equatorial wave responsible for producing the well known 5-micron hotspots observed at infrared wavelengths. Plume $P1$ can be identified in both the top and bottom optical panels and its trajectory slightly leads a radio hotspot in the middle radio panel. We also identify $P2$ as an optical plume feature associated to a different radio hotspot after a similar parallel trajectory (parallel to $P1$'s trajectory), is projected between the top and middle panels. The plumes appear to have a small convective head that produces the high altitude, white clouds that experience horizontal shear and fan out over time. This proposed vertical and horizontal motion depletes the surrounding atmosphere of ammonia as the plumes migrate eastward, and the radio hotspots appear as oval shaped regions located just northward of the plumes matching numerical models \citep{2000_Showman, 2002_Baines, 2005_Friedson}

\begin{sidewaysfigure}
\begin{center}
\includegraphics[scale=0.7]{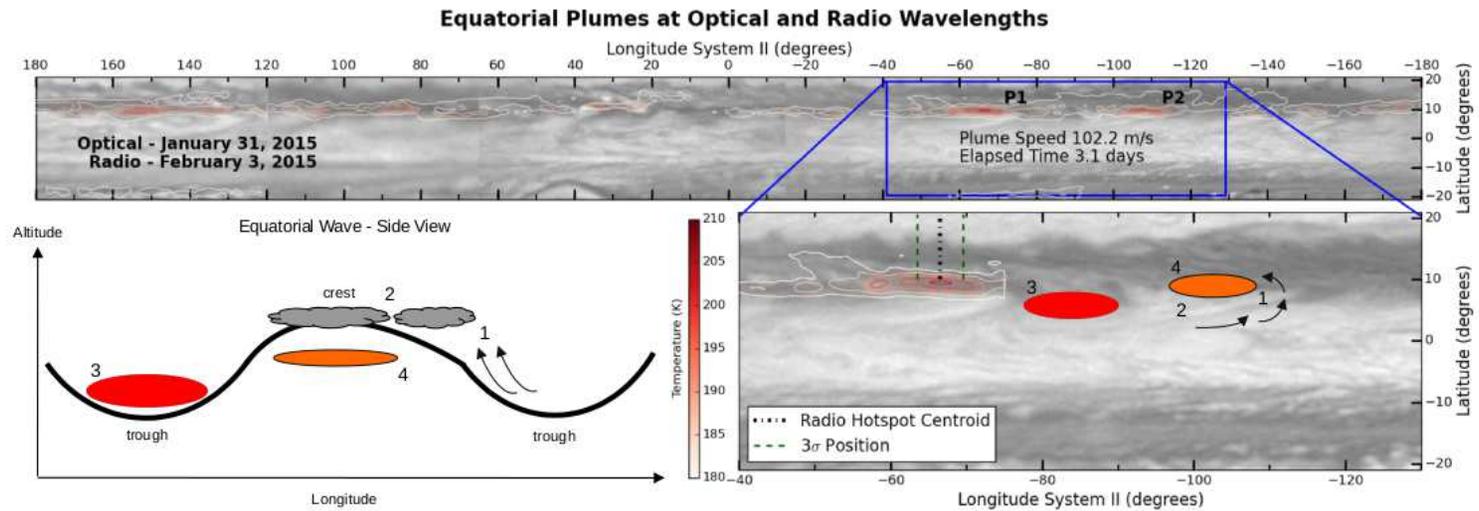}
\caption{Top panel in Figure \ref{VLA_EQ} is shown with overlaid radio intensity contours. Plumes P1 and P2 are identified again and their surrounding area is enlarged in the lower right panel. The temperature colorbar applies to the radio contours in both optical panels. The centroid of the radio hotspot associated with P1 and its 3$\sigma$ position uncertainties are located. The schematic in the lower left panel represents the morphology of the equatorial wave, optical white plumes, optical dark spots between the plumes, 5-micron hotspots observed in infrared, and the radio hotspots observed at 2 cm. Features 1-4 are: ``1" - convective plume head, ``2" - high altitude ammonia rich clouds, ``3" - 5-micron infrared hotspot (red and hatched), and ``4" - 2 cm radio hotspot (orange). Arrows show motion associated with the plumes according to \citet{2013_Choi}. \label{plume}}
\end{center}
\end{sidewaysfigure}

We present the top panel in Figure \ref{VLA_EQ} again but this time with overlaid radio intensity contours in the top panel of Figure \ref{plume}. The dates for the observations and wavelengths are shown on the left. We used the wave velocity at this latitude from \citet{2016_dePater} to longitudinally translate the radio contours back approximately 3.1 days to coincide with the optical mosaic. We limited the range of brightness temperatures from the observation to explore the radio hotspots' morphology in greater detail and the color bar in Figure \ref{plume} applies to the radio contours in both optical panels. The lower right panel of Figure \ref{plume} is an enlarged plume region, where the radio contours associated with $P1$ are shown. We suppressed the contours near $P2$ in favor of identifying numbered features corresponding to the schematic in the lower left of Figure \ref{plume}. 

The lower left panel in Figure \ref{plume} depicts a side view of the equatorial wave and the region's hypothetical morphology with limited dynamics. The thick wave black line denotes the equatorial wave with a crest and two troughs. Wave induced up-welling occurs at ``1" where cyclonic motion has been observed to correspond to a convective plume head \citep{2013_Choi}. High altitude ammonia rich clouds ``2" are produced that eventually experience shear and fan out to form the familiar white plume shapes, like P1 and P2. The infrared hotspots, ``3" are observed between these plumes, corresponding to optical dark patches and likely reside in the troughs of the wave \citep{1996_Orton}. The radio hotspots ``4" are not associated with the optical dark patches, and therefore, are not associated with the infrared hotspots and appear more connected to the dynamics of the plume. One proposition is that the wave induced up-welling at ``1" depletes a northward region below the plume clouds, at a pressure range sensitive to this radio observation wavelength, of ammonia, which we observe as radio hotspots ``4". That ammonia rich gas is sent transported to produce the high altitude clouds plume fan clouds, ``2". As the wave and plume head move, the depleted ammonia region is expanded longitudinally.      

To prove that the radio hotspot can not be confused with the infrared hotspots, we located the centroid of the radio hotspot associated with $P1$ and found its latitude to be $\sim9.5^\circ$N and longitude, in system II coordinates, to be $\sim-66.5^\circ$W. We used the velocity uncertainty reported in \citet{2016_dePater}, $\pm1.4\rm ms^{-1}$, and the VLA observation resolution, $1.4^\circ$, to determine the $3\sigma$ uncertainty in the centroid position of this radio hotspot. We clearly see that its position likelihood is located north of plume $P1$ in latitude and longitudinally aligned with $P1$ rather than an optical dark patch between adjacent plumes, while the infrared hotspots align with the optically dark patches \citep{1996_Orton}. Even with larger uncertainties for the wave velocity, the VLA's resolution for this observation distinctly separates the radio hotspots from the infrared hotspots in latitude and longitude. 

\section{Discussion and Conclusions}

The upcoming Juno mission to Jupiter aims to probe the planet's deep atmosphere ($>$100s bars) through radio wavelength observations by the spacecraft as it periodically orbits interior to the radiation belts that hinder similar Earth based observations. The spacecraft will not remain there indefinitely though, and will undergo a steady diet of radiation poisoning over the mission's lifetime and eventually be sent on a final mission-ending plunge into Jupiter's atmosphere where it will ultimately be crushed, melted and become a part of Jupiter itself. Although Juno will provide valuable information about Jupiter's atmosphere and interior, Earth based observations will prove crucial in understanding the connection between the pressures observed during the Juno mission to the layers accessible at wide range of pressures with facilities like HST, VLA, the NASA Infrared Telescope Facility (IRTF) atop Mauna Kea, and the Atacama Large Millimeter Array (ALMA). We utilized our VLA observation to maximize its scientific information through a variety of analyses that may open new doors to complement Juno and future like missions and observations. 

Most relevant to the Juno mission's deep circulation science goal is our spectral analysis of the radio map shown in Figure \ref{fig1}. The retrieved spectra indicate most likely that Jupiter's atmosphere accessed at 2 cm is in a 2-D turbulent regime and is governed by dynamics of the shallow water equations. In the 2-D turbulent regime, energy is pumped into the shallow atmosphere at small scales and is inversely transferred to larger scale features, such as the GRS and the large planetary equatorial wave associated with the infrared and radio hotspots. When latitudes encompassing the radio hotspots are excluded, the passive tracer power spectrum from Figure \ref{fig1} has slopes of $m_1 = -1.6$ and $m_2 = -1.0$ which are in agreement with theoretical spectral indices for a 2-D turbulent atmosphere. The forcing scale was identified at $k_f = 27$ which translates to a physical size near the Rhines scale $\sim$13,000 km at the equator. It is also remarkably similar to the wavenumber transition identified in \citet{1996_Harrington} and should be investigated further considering this forcing scale connects two different regions of Jupiter's atmosphere.

The bright radio-hot belt appears to be connected to the 5-micron hotspots and the equatorial wave at their respective latitudes. We identified the radio hotspots' primary wavenumber at $k = 7$ in Figure \ref{fig2D_wave} which is consistent with past observations of $k = 8$. However, the secondary wavenumber of $k = 17$ has not been previously mentioned in findings at optical or infrared wavelengths. Follow up investigations at multiple wavelengths could reveal that the hotspots features are just as dynamic below the visible clouds \citep{2010_dePater}. Our results suggest that the equatorial wave is either a superposition of two waves or that the spots split and coalesce again, repeatedly, as seen in optical studies \citep{2013_Choi}.

We have shown that Jupiter's banded appearance in our radio observation supports the model that bright, optical wavelength zones are regions of ammonia rich gas, consistent with multi-frequency radio map analyses \citep{2016_dePater}. Meanwhile, the dark belts, appear as radio bright regions where ammonia is depleted and emission originates from warmer layers at higher pressures in Jupiter's atmosphere (see \citet{2016_dePater} for radiative transfer calculations). The strong correlations connect the zonal wind shear, $U_y$, and radio banding for selected latitudinal regions in the southern hemisphere, while the optical wavelength banding and radio intensity seem to match only near the Jupiter's equatorial region. To prompt further studies of these correlations, one ideally would utilize higher radio frequencies to increase resolution and lower frequencies to acquire profiles at larger pressure ranges. The most obvious and interesting finding regarding these profiles is the asymmetry we find between the Jovian northern and southern hemispheres. The southern hemisphere, excluding GRS latitudes, shows correlations of $U_y$ to the radio profile, but the same behavior is not observed in the northern hemisphere. Could the radio profile change over time and indicate some dynamical difference between the northern and southern hemispheres? These tantalizing questions may be answered soon enough from Juno mission findings.  

We compared our VLA data to low resolution amateur images and found corresponding features observed at different wavelengths that are vertically connected. Although such analysis for observations separated by several days was successful in tracking features qualitatively, more accurate quantitative analysis in future studies are desired. Where we lacked in precision however, we proved in concept, and new questions have arisen about Jupiter's equatorial region such as, what are the radio hotspots? How are they connected to the equatorial wave? How are they connected to the 5-micron hotspots and the optical bright plumes? We have shown that some radio hotspots in our observation reside slightly northward of the optical plumes and do not align in longitude or latitude to the optical dark patches between adjacent plumes like the 5-micron infrared hotspots \citep{1996_Orton, 2016_dePater}. But, we also find radio hotspots not associated with any optical plume, and additionally, there are optical regions containing bright clouds with no accompanying radio hotspot. The dynamics of the NEB could be coupled to the appearance of radio hotspots more so than the infrared hotspots but only more simultaneous multi-wavelength observations will begin to shed light on this problem.  

The long lived nature of the 5-micron hotspots and appearance of other small scale radio features seen in Figure \ref{fig1} give credence that feature tracking techniques implemented at optical wavelengths could be applied to multiple radio wavelength maps and directly measure motion in Jupiter's atmosphere below the upper cloud layer. A single map can be used to infer the dynamics, as mentioned earlier, in our wave power spectral analysis, but one ideally desires to track the motions of small scale radio features and be the first to directly measure a radio derived observational zonal wind profile. A zonal wind profile below the upper cloud deck, spanning $45^\circ$N to $45^\circ$S, could obtain a precision of $14\rm ms^{-1}$ for three maps produced from consecutive nights of observing with the VLA in the B-configuration. The strong equatorial jets and the strong eastward jet at $23^\circ$N represent excellent latitudes to compare observed speeds at radio pressure ranges to HST cloud deck velocities and address a long standing question of how the eastward jet velocities change with altitude. The same is true for weaker westward jets away from the equator where the three consecutive maps would prove more vital in measuring their lower velocities. Zonal wind profiles from HST and the VLA could then be compared to theoretical works \citet{1989_Dowling, 2004_Gierasch} and modeling studies of individual jets' behavior \citet{2005_Garcia, 2008_SanchezLavega}. Vertical wind structure as a function of latitude for pressures greater than 0.7 bars is still currently a free parameter in giant planet GCMs because of the lack of observations. The Juno mission will be able to indirectly estimate motion at pressures between near 100 bars, but motion observed at 2 cm (VLA) and 3 mm (ALMA) could serve as observational data to constrain the free parameter of vertical wind shear in GCMs.

Simultaneous observations with HST and the VLA could also reveal information about the vertical motion of ammonia in Jupiter's atmosphere. The map in Figure \ref{fig1} and close up in Figure \ref{fig-vis-rad} show strong correlations to the vertical extent of spots, but only at a particular instance of time. If simultaneous maps are taken over consecutive days, the time variability of regions near spots and oval BA could yield observations of SIAC's in the optical that correspond to radio counterparts of regions with high ammonia where vertical motion and shear could be investigated in detail. Perhaps the life cycle of a convective region could be detailed in multiple stages; where first an area of increased ammonia is observed in the radio, followed by optical SIAC's in the same latitude and longitude region but vertically higher afterwards, and eventually the appearance of a depleted region, like the RH in Figure \ref{fig-vis-rad}, seen again at radio wavelengths. 

Coordinating complex and long duration observations with multiple telescopes is a highly expensive endeavor but the simultaneous data at multiple levels of the atmosphere would be invaluable. The Juno mission has seen a cooperative multi-wavelength ground based support observation campaign and stands to provide an unprecedented look at Jupiter's atmosphere.  

%\end{linenumbers}

\section*{Acknowledgement}
The National Radio Astronomy Observatory is a facility of the National Science Foundation operated under cooperative agreement by Associated Universities, Inc. Richard Cosentino is a student at the National Radio Astronomy Observatory. Support for this work was provided by the NSF through the Grote Reber Fellowship Program administered by Associated Universities, Inc./National Radio Astronomy Observatory.

\biboptions{authoryear}

\bibliographystyle{apj}
\bibliography{bibliography}
 
\clearpage

\clearpage

\clearpage

\clearpage

\clearpage

\clearpage

\clearpage

\clearpage

\clearpage

\clearpage

\clearpage

\begin{table}
\begin{center}
\textbf{Table 1} \\
\textbf{Power Law Fits to Jovian Jets} \vspace{1mm} \\
\renewcommand{\arraystretch}{1}
\begin{tabular}{lllcccc}
\hline 
\hline
\centering
Author & Region & Latitude & Wavenumber & Slope Fit & Wavenumber & Slope Fit \\
\centering
Year & Name & degrees & $k$ & $m$ & $k$ & $m$ \\ \hline 
\\
%\multicolumn{5}{l}{Harrington et. al. 1996} \\
%\cline{1-6}
\centering
Harrington & Eastward Jets & Various & 1-24 & -0.720 $\pm$ 0.001 & 28-60 & -2.709 $\pm$ 0.07 \vspace{2mm} \\
et al. 1996 & Westward Jets & Various & 1-24 & -0.715 $\pm$ 0.001 & 28-60 & -3.137 $\pm$ 0.12 \vspace{2mm} \\ 
%\multicolumn{5}{l}{This Work} \\ 
\cline{1-7} \\
Cosentino & Eastward Jets & Various & 5-24 & -1.1 $\pm$ 0.2 & 28-60 & -1.9 $\pm$ 0.4 \vspace{2mm} \\
at al. 2016 & & & & & 28-114 & -1.6 $\pm$ 0.1 \vspace{2mm} \\
& Westward Jets & Various & 5-24 & -0.9 $\pm$ 0.2 & 28-60 & -1.7 $\pm$ 0.6 \vspace{2mm} \\
 & & & & & 28-114 & -0.9 $\pm$ 0.2 \vspace{2mm} \\
\hline
\end{tabular}
\end{center}
\caption{This table format was based on \citep{1996_Harrington} and updated to include this work. \label{table1}}
\end{table}

\clearpage

\begin{table}
\begin{center}
\textbf{Table 2} \\
\textbf{Composite Model Fits to Regions of Spectra} \vspace{1mm} \\
\renewcommand{\arraystretch}{1}
\begin{tabular}{llccc}
\hline 
\hline
\centering
Region & Latitude & Slope 1 & Slope 2 & Forcing Scale  \\
Name & degrees & $m_1$ & $m_2$ & $k_f$ \\ \hline 
Eastward Jets & Various & -0.8 $\pm$ 0.2  & -1.4 $\pm$ 0.1 & 14 $\pm$ 2 for $k_{mod}\leq$9 \vspace{2mm} \\
Westward Jets & Various & -1.0 $\pm$ 0.2 & -0.8 $\pm$ 0.1 & 36 $\pm$ 5 for $k_{mod}\leq$17 \vspace{2mm} \\
Entire Map & 60$^\circ$S - 60$^\circ$N & -1.9 $\pm$ 0.3 & -1.2 $\pm$ 0.1 & 13 $\pm$ 1 for $k_{mod}\leq$8 \vspace{2mm} \\
Hotspots Region & 8$^\circ$N - 12$^\circ$N & 0.0 $\pm$ 0.6 & -1.2 $\pm$ 0.2 & 18 $\pm$ 2 for $k_{mod}\leq$13 \vspace{2mm} \\
Map Excluding Hotspots Region & 60$^\circ$S - 8$^\circ$N & -1.6 $\pm$ 0.2 & -1.0 $\pm$ 0.1 & 27 $\pm$ 5 for $k_{mod}>$9 \\
 & 12$^\circ$N - 60$^\circ$N  & & & \\
\hline
\end{tabular}
\end{center}
\caption{The convergence of fits for different test spectra model lengths, $k_{mod}$, is presented in the last column. \label{table2}}
\end{table}

\clearpage

\begin{table}
\begin{center}
\textbf{Table 3} \\
\textbf{Regional Analysis - Correlation Coefficients and p-values} \vspace{1mm} \\
\renewcommand{\arraystretch}{1}
\begin{tabular}{llccc}
\hline 
\hline
\centering
Profiles & Region & Latitudes & Correlation & P-value   \\
 &  &  & Coefficient &  \\ \hline 
Winds U and Radio & A & 36.7$^\circ$S - 51.3$^\circ$S & 0.31 & 0.01 \vspace{2mm} \\
Winds U and Radio & A (ext) & 34.9$^\circ$S - 55.6$^\circ$S & 0.19 & 0.07 \vspace{2mm} \\
U${_y}$ and Radio & A & 36.7$^\circ$S - 51.3$^\circ$S & 0.77 & 8e-14 \vspace{2mm} \\
U${_y}$ and Radio & A (ext) & 34.9$^\circ$S - 55.6$^\circ$S & 0.56 & 8e-9 \vspace{2mm} \\
Winds U and Radio & B & 0.5$^\circ$S - 14.6$^\circ$S & 0.00 & 0.99 \vspace{2mm} \\
U${_y}$ and Radio & B & 0.5$^\circ$S - 14.6$^\circ$S & 0.76 & 4e-13 \vspace{2mm} \\
658 nm and Radio & C & 35.1$^\circ$S - 35.1$^\circ$N & -0.49 & 7e-20 \vspace{2mm} \\
\hline
\end{tabular}
\end{center}
\caption{The latitude bounds of Region A where extended to Region A extended ``(ext)" to explore analysis senstivity. \label{table3}}
\end{table}

% The Appendices part is started with the command \appendix;
% appendix sections are then done as normal sections
%\appendix

%% Using an acknowledgements command is not in the Elsevier template,
%% but it can be used.
%\ack
%This work has made use of NASA's Astrophysics Data System.  It 
%also benefitted tremendously from \citet{latexguide}.

\label{lastpage}

% Bibliographic references with the natbib package:
% Parenthetical: \citep{Bai92} produces (Bailyn 1992).
% Textual: \citet{Bai95} produces Bailyn et al. (1995).
% An affix and part of a reference:
%   \citep[e.g.][Ch. 2]{Bar76}
%   produces (e.g. Barnes et al. 1976, Ch. 2).-

%\bibliography{bibliography.bib}
%\bibliographystyle{elsart-harv}

%\bibliography{cumulative.bib}

%%%% Use the plainnat style for ``Icarus'' mode to display DOI numbers
%%%% among other things.  However, revert to the Elsevier elsart-harv
%%%% mode for ``Elsevier'' mode.

%\bibliography{mybibfile}
%\bibliographystyle{plainnat}

%\bibliography{bibliography}

%\bibliographystyle{apj2}

\end{document}